\documentclass{iopart}

\usepackage{mathrsfs}

\usepackage{amssymb}

\usepackage{verbatim}

\begin{document}

\newcommand{\la}{\label}
\newcommand{\be}{\begin{eqnarray}}
\newcommand{\ee}{\end{eqnarray}}
\newcommand{\der}{\partial}
\newcommand{\w}{\tilde}
\newcommand{\str}{\,{\mathrm{str}}}
\newcommand{\sdet}{\,{\mathrm{sdet}}}
\newcommand{\re}{{\mathrm{Re}}\,}
\newcommand{\im}{{\mathrm{Im}}\,}
\newcommand{\Li}{\,\mathrm{Li}\,}
\newcommand{\m}{\mathrm{m}}
\newcommand{\diag}{\mathrm{diag}}
\newcommand{\bor}{{\bf r}}
\title[Universal and non-universal features of the multifractality exponents.]{Universal and non-universal features of the multifractality exponents of critical wave-functions.}

\author{I. Rushkin, A. Ossipov, Y. V. Fyodorov}
\address{School of Mathematical Sciences, University of Nottingham, Nottingham NG7 2RD, United Kingdom}

\date{\today}

\begin{abstract}
We calculate perturbatively the multifractality spectrum of wave-functions in critical random matrix ensembles in the regime of weak multifractality. We show that in the leading order the spectrum is universal, while the higher order corrections are model-specific. Explicit results for the anomalous dimensions are derived in the power-law and ultrametric random matrix ensembles.

\end{abstract}



\maketitle
\section{Introduction}

Critical wave-functions $\psi(\bor)$ at the point of the Anderson localization transition \cite{EM08} are usually\footnote{For a rare  exception to this rule see \cite {ORC}.}  multifractal \cite{W80}. Their statistical properties can be characterized by a set of exponents $d_q$, which describe the scaling of the wave-function moments with the system size $L$ in the limit $L\to \infty$. Using the bar to denote the disorder averaging, one can define $d_q$ via the relations
\be
\la{formula number 1}
I_q=\int d^d \bor \, \overline{|\psi(\bor)|^{2q}}\propto L^{-d_q(q-1)}\,.
\ee
For completely extended states $d_q=d$; for completely localized states $d_q=0$. The hallmark of multifractality is
a $q$-dependent $d_q$, different from both $0$ and $d$.

Although detailed characterization of the critical wave-function multifractality remains an active area of research (see \cite{EM08}), relatively few explicit analytical results for $d_q$ are available in the literature. One such result is due to Wegner \cite{W86}, who calculated the anomalous multifractal dimensions $\Delta_q=(d_q-d)(q-1)$ for the Anderson transition in $2+\epsilon$ dimensions:
\be\label{Wegner}
\Delta_q&=&q(1-q)\epsilon+\frac{\zeta(3)}{4}q(q-1)(q^2-q+1)\epsilon^4+O(\epsilon^5),\quad\beta=1,\nonumber\\
\Delta_q&=&q(1-q)\left(\frac{\epsilon}{2}\right)^{1/2}-\frac{3\zeta(3)}{8}q^2(q-1)^2\epsilon^2+O(\epsilon^{5/2}),\quad\beta=2,
\ee
with $\zeta(x)$ standing for the Riemann zeta-function.
These expansions were derived in the replicated nonlinear $\sigma$-model with a
coupling constant $t\propto\epsilon$ in the orthogonal symmetry class (time-reversal invariant
systems, $\beta=1$) and $t\propto\sqrt{\epsilon}$ in the unitary symmetry class (systems with broken time-reversal invariance, $\beta=2$). Note that the second and third order corrections are equal to zero in both cases.

Another explicit result was derived more recently \cite{ME00,KOYC10} in the framework of
power-law banded random matrices (PLBRM), introduced in \cite{MFDQS1996} as a useful toy-model with multifractal eigenstates. In the regime of weak multifractality the anomalous exponents are given in the lowest order by\footnote{In this work $t$ is $8\pi$ times smaller than in the cited reference.}
\be\label{PLBRM}
\Delta_q=q(1-q)\frac{t}{\beta}+O(t^2),
\ee
where $t/\beta$ plays the role of the coupling constant in the effective $\sigma$-model and is assumed to be small. Comparing eq. (\ref{Wegner}) with eq. (\ref{PLBRM}) we notice that the leading orders are the same if one identifies $\epsilon$ ($\sqrt{\epsilon/2}$ for $\beta=2)$ with the coupling constant $t/\beta$ of the corresponding $\sigma$-model. This observation can be generalized to an arbitrary model at criticality in the regime of weak multifractality (see Section \ref{section-sigma}), making the leading order contribution in $\Delta_q$ model-independent or {\it universal}. While the critical exponents of short-ranged models are numbers, in the long-ranged models they are functions of parameters (e.g. of bandwidth in PLBRM). Universality then means that the functional forms of critical exponents from different models can be mapped onto each other. Remarkably, in PLBRM, a similar universality also exists in the opposite regime of strong multifractality \cite{FOR09}.
\mathchardef\mhyphen="2D

The aforementioned universality poses a question: does the knowledge of higher order contributions to $\Delta_q$ allow to distinguish among different models? These contributions may contain model-specific information, making $\Delta_q$ different for different models within the same symmetry class. One should stress, however, that since $\Delta_q$ is a critical exponent, it is expected to be insensitive to any short-scale deformation of a model within its universality class.

In the present work we calculate the anomalous multifractal dimensions to the second order in the coupling constant for two critical random matrix ensembles: PLBRM and the ultrametric random matrices \cite{FOR09}. In the unitary symmetry class ($\beta=2$), we find that the second order correction is equal to zero. It follows from our derivation that the absence of the second correction is very general and not restricted to the particular models we consider. In the orthogonal case ($\beta=1$), the second order correction turns out to be non-zero and different for the two models considered. This fact should be contrasted with the situation in $2+\epsilon$ dimensions
and may be a consequence of the long-range nature of the model Hamiltonian. In addition, we find that the first deviation of the multifractal spectrum from its parabolic shape occurs in the third order correction and is also model-specific.

\section{Perturbation theory in the non-local non-linear $\sigma$-model}\label{section-sigma}
Non-linear supersymmetric $\sigma$-model is the standard field-theoretical description of a single quantum particle in a disordered medium \cite{Efetov book}. When computing critical exponents, insensitive to short-scale details, one can take the equivalent $\sigma$-model description as the starting point instead of using the underlying random matrix-type Hamiltonian explicitly.  From this angle, let us consider a 1D chain of $L$ sites enumerated by a discrete spatial coordinate $r$, and associate with every site a supermatrix $Q_r$. The effective $\sigma$-model is defined by its action:
\be\la{0} S[Q]= -\frac{\beta}{16t}\sum_{r,r'}U_{r,r'}\str (Q_rQ_{r'})-i\eta\sum_r\str(Q_r\Lambda).\ee
Here the kernel $U_{r,r'}$ defines the non-local term in the action (``kinetic energy"), and can be viewed as a symmetric $L
\times L$ matrix with a zero-mode: $\sum_{r}U_{r,r'}=0$. The coupling constant $t\ll 1$ is the small parameter of the perturbative expansion, and $\beta$ is the usual symmetry parameter of the random matrix theory: $\beta=1$ in the orthogonal symmetry class ($Q$ being a $8\times8$ matrix) and $\beta=2$ in the unitary symmetry class ($Q$ being $4\times4$). The matrix $\Lambda=\diag(\mathbf{1},-\mathbf{1})$ in the advanced-retarded representation, and $\eta>0$ represents the effective level broadening.

Let us sketch the main steps of the standard perturbative treatment of the model (see \cite{FM1995,KM94} and references therein for more details).
The wave-function moments (\ref{formula number 1}) (also known as the inverse participation ratios or  IPR's) can be extracted as
\be\la{1} I_q=\lim_{\eta\to 0} \frac{\eta^{q-1}}{4L}\int \mathcal{D}Q e^{F[Q]},\quad e^{F[Q]}=\Bigl(\sum_r \str^q(Q_r\Lambda k)\Bigr)e^{-S},\ee where $ k=\mathbf{1}\otimes\diag(1,-1)$ in the advanced-retarded representation.
Following \cite{FM1995}, we will use the rational parametrization of the matrix $Q(r)$ with the zero-mode $Q_0$ separated:
\be\fl\qquad Q_r=T\Lambda g(W_r)T^{-1},\quad Q_0=T\Lambda T^{-1}, \quad g(W)=(1+\frac{1}{2}W)(1-\frac{1}{2}W)^{-1}.\ee
Here $T$ is a position-independent matrix and the spatial variations are represented by $W_r$ -- a small ( $O(\sqrt{t})$) block-off-diagonal matrix with a zero mode: $\sum_rW_r=0$.
The change of variables $Q\to\{Q_0,W\}$ in the functional integral incurs a Jacobian, known to be $J=\exp\Bigl[\frac{1}{8L}\sum_r W_r^2\Bigr]+O(t^2)$ in the orthogonal case, and $O(t^2)$ in the unitary case. The perturbation series is based on the Taylor expansion $g(x)=1+x+\frac{1}{2}x^2+...$ Substituting such an expansion into the action (\ref{0}), one separates the $O(1)$ part of the ``kinetic energy": $S_0=\frac{\beta}{16t}\sum_{r,r'} U_{r,r'}\str(W_rW_{r'})$ and treats all the other terms in (\ref{1}) as a perturbation. The averages $\langle\cdots\rangle$ with respect to $S_0$ are computed by the Wick theorem based on the elementary contractions:
\be\fl \beta=2:\quad &\langle \str(W_rP)\str(W_{r'}R)\rangle=t\Pi_{r,r'}\str(PR-P\Lambda R\Lambda),\nonumber\\
&\langle\str(W_rP W_{r'}R)\rangle=t\Pi_{r,r'}\Bigl(\str P\str R - \str(P\Lambda)\str(R\Lambda)\Bigr),\ee
\be\fl &\beta=1:\quad \langle \str(W_rP)\str(W_{r'}R)\rangle=t\Pi_{r,r'}\str(PR-P\Lambda R\Lambda+P\Lambda\w R\Lambda -P\w R),\nonumber\\
\fl&\langle\str(W_rP W_{r'}R)\rangle=t\Pi_{r,r'}\Bigl(\str P\str R - \str(P\Lambda)\str(R\Lambda) +\str(P\Lambda\w R\Lambda -P\w R)\Bigr).\ee
Here $P$ and $R$ are arbitrary matrices and $\w R=M^TR^TM$, where $M$ is an orthogonal matrix of the time-reversal transformation. The propagator $\Pi_{r,r'}$, associated with the action $S_0$, is defined by the identity
$\sum_r U_{r_1,r}\Pi_{r,r_2} = \delta_{r_1,r_2} -\frac{1}{L}$,
where the subtraction is required by the zero-mode of $U_{r,r'}$; it also ensures that the propagator has a zero-mode as well: $\sum_r\Pi_{r,r'}=0$. Moreover, we consider spatially uniform systems in which the diagonal elements $U_{r,r}$ and $\Pi_{r,r}$ are independent of $r$.
Implementing such a perturbation theory, one effectively integrates out all the factors $W_r$ containing non-zero momenta, and is left with the integral over the zero-mode $Q_0$ only,
which is simplified in the limit $\eta\to 0$ and can be done using any standard parametrization of this matrix \cite{Efetov book}. As a result, one can prove in the unitary case the following correspondence: given that
\be\fl\quad e^{F_\mathrm{eff}[Q_0]}=L\str^q(Q_0\Lambda k)\Bigl[A_0 + A_1\Bigl(\epsilon L\str(Q_0\Lambda)\Bigr)+ \ldots +A_n\Bigl(\epsilon L\str(Q_0\Lambda)\Bigr)^n\Bigr],\ee
where $A_i$ are some constants (in the present calculation only the first three are non-zero), the zero mode integrals over $Q_0$ yield
\be\fl\la{zero-mode integral}\quad I_q=\frac{q!}{L^{q-1}}\Bigl(A_0+ A_1(q-1) +\ldots +A_n\frac{(q+n-2)!}{(q-2)!}\Bigr),\ee
and a similar expression with the prefactor $(2q-1)!!/L^{q-1}$ in the orthogonal case.

Performing a systematic expansion up to the order $t^2$ in this way, we find for the unitary case:
\be\fl\la{unitary expansion} I^{(U)}_q=\frac{q!}{L^{q-1}}\left(1 \!+\!t\frac{q(q\!-\!1)}{2}\Pi_{r,r}\!+\!t^2\frac{q^2(q\!-\!1)^2}{8}\Pi^2_{r,r}\!-\!t^2\frac{q(q\!-\!1)(2q\!-\!1)}{4}\Pi_2\right)\! +\!O(t^3),\ee
where $\Pi_2=L^{-2}\sum_{r,r'}\Pi^2_{r,r'}$.
To obtain this result, one needs to keep track of terms with up to four $W_r$-factors everywhere except in the expansion of the ``kinetic energy" in $S$, where one has to go up to six $W_r$'s because of the $1/t$ prefactor. Remarkably, in the unitary case these extra terms vanish after averaging over $W_r$, as does the $O(t^2)$ contribution of the Jacobian $J$. The latter statement is still true in the orthogonal case, but the former is not: due to a more complicated structure of the $W\mhyphen W$ contractions in this case, a ``kinetic" contribution $K$ survives the average:
\be\fl\quad\la{orthogonal expansion}I^{(O)}_q=\frac{(2q\!-\!1)!!}{L^{q-1}}\!\left(1 +tq(q\!-\!1)\Pi_{r,r}+t^2\frac{q^2(q\!-\!1)^2}{2}\Pi^2_{r,r}+t^2\frac{q(q\!-\!1)}{2}\Pi^2_{r,r}-\right.\nonumber\\\left.t^2\frac{q(q\!-\!1)(4q\!-\!5)}{2}\Pi_2+t^2q(q-1)K\right) +O(t^3),\ee
\be\fl\la{kinetic term} \quad K=L^{-1}\sum_{r,r',r''}U_{r,r'}\Pi_{r,r'}\Pi_{r,r''}\Pi_{r',r''}-\Pi^2_{r,r}-(L-1)\Pi_2.\ee
Also, unlike the unitary case, the Jacobian $J$ has an $O(t)$ part that contributes to the penultimate term in the brackets of (\ref{orthogonal expansion}).

In the limit of weak multifractality, the anomalous dimensions $\Delta_q$ are parametrically small and hence $I_q\propto L^{1-q}\exp(-\Delta_q\log L)= L^{1-q}\Bigl(1-\Delta_q\log L +\frac{1}{2}\Delta_q^2\log^2L+O(\Delta_q^3))$. Having found the large $L$ asymptotic behavior of equations (\ref{unitary expansion}) and (\ref{orthogonal expansion}), we can check that it is indeed of the required form and extract  $\Delta_q$ up to the order $t^2$. Note that in order to demonstrate the first proposition it is sufficient to show that for large $L$
(i) the kinetic contribution behaves as $K \to -\frac{1}{2}\Pi^2_{r,r}+c_1\log L + c_2$, and (ii) $\Pi_2$ tends to a constant. The latter appears to be a universal fact: if $\Pi_2$ grew with $L$, it would contribute to the anomalous dimension. However, the prefactor in front of $\Pi_2$ does not have the symmetry $q\to 1-q$, which the anomalous dimension of any $\sigma$-model should possess on general grounds \cite{MFME2006}. Therefore, we can see already from eq. (\ref{unitary expansion}) that in the unitary case the second order correction to the anomalous dimension vanishes. Below we will apply the results of this section to two types of critical systems: the $\sigma$-model corresponding to the critical
PLBRM ensemble \cite{MFDQS1996}  as well as the critical $\sigma$-model, which can be derived from the $n$-orbital version of the ensemble of ultrametric random matrices introduced in \cite{FOR09}.

\section{Critical Power-law Banded Random Matrix ensemble}
Let us briefly recall the definition of the critical PLBRM and the corresponding $\sigma$-model. Consider a system of $L$ sites  whose Hamiltonian is an $L\times L$ random matrix $H_{r,r'}=G_{r,r'}a(|r-r'|)$, and further confine it
to a circle by imposing periodic boundary conditions. The random matrix $G$ is taken from the standard Gaussian ensemble (unitary or orthogonal), and $a(r)$ is a function that controls the spatial decay of the disorder: $a(r)\simeq 1$ for $r\lesssim b$ and $a(r)=b/r$ as $r\to\infty$. In the limit of large bandwidth $b\gg 1$ this model can be mapped \cite{MFDQS1996} onto the non-local $\sigma$-model (\ref{0}) with
\be\la{definitions} U_{r,r'}=(A^{-1})_{r,r'}-\delta_{r,r'}\Bigl(\sum_i A_{i,j}\Bigr)^{-1}, \quad t=\frac{1}{2\pi b}\ll 1,\ee
where we introduced an $L\times L$ matrix $A$ with entries $A_{r,r'}=a^2(|r-r'|)$. The purpose of subtracting the diagonal term in $U_{r,r'}$ is to ensure $\sum_rU_{r,r'}=0$. Translational invariance suggests working in the momentum space. For small momenta $k\ll L$, the discrete nature of the underlying lattice becomes irrelevant, and we can safely employ the
long-wave asymptotic of the Fourier components: $\w{U}_k\sim |k|$ and $\w\Pi_k\sim 1/|k|$. In fact, since the anomalous dimensions must be insensitive to short-scale details of the model, only such small momenta are important and one can use the above asymptotic expressions as exact definitions for an effective model. Defining the cutoff $k_\m$ via $L=2k_\m+1$, we then arrive at:
\be\fl\quad U_{r,r'}=2L^{-2}\sum_{k=\pm1,\ldots}^{|k|=k_\m}e^{i\frac{2\pi}{L}k(r-r')}|k|,\quad \Pi_{r,r'}=\frac{1}{2}\sum_{k=\pm1,\ldots}^{|k|=k_\m}e^{i\frac{2\pi}{L}k(r-r')}\frac{1}{|k|}.\ee
In particular, $\Pi_{r,r}=\sum_{k=1}^{k_\m}1/k=\log(k_\m e^\gamma) +O(1/k_\m)$, where $\gamma\approx0.577$ is the Euler-Mascheroni constant, and $\Pi_2=\frac{1}{2}\sum_{k=1}^{k_\m}1/k^2=\frac{\pi^2}{12} +O(1/k_\m)$. These formulas are sufficient for our purpose in the unitary case:
\be\fl\quad\la{preliminary1unit}I^{(U)}_q=\frac{q!}{L^{q-1}}\left( 1+t\frac{q(q-1)}{2}\log L +t^2\frac{q^2(q-1)^2}{8}\log^2L + t^2c\right) +O(t^3),\ee
where only the leading terms in $L\to\infty$ are kept and by $c$ we hereafter denote a constant independent of both $L$ and $t$. The above expression implies
\be \la{answer1unit}\Delta_q=\frac{q(1-q)}{2}t +O(t^3),\quad \beta=2.\ee
The second order correction is zero, as announced. The $\beta=1$ case is less trivial, since eq. (\ref{kinetic term}) also requires the asymptotic \be\la{term to compute} \fl\quad\sum_{r,r',r''}U_{r,r'}\Pi_{r,r'}\Pi_{r,r''}\Pi_{r',r''}=L\!\Bigl(\frac{\pi^2}{6} k_\m+\frac{1}{2}\Pi_{r,r}^2 -\Pi_{r,r} +c\Bigr) +O(k_\m^{-1}),\ee
which is extracted using the Euler-Maclaurin formula \cite{AS1972} and the asymptotic expansion of the digamma function at infinity.
Note that the leading term $\frac{\pi^2}{6}k_\m \sim \frac{\pi^2}{12}L$ (which would be incompatible with the assumption of multifractality) cancels out with the last term in eq. (\ref{kinetic term}). As a result,
\be\fl\la{preliminary1orth}I^{(O)}_q=\frac{(2q\!-\!1)!!}{L^{q-1}}\Bigl(1 +t(1-t)q(q-1)\log L + t^2\frac{q^2(q-1)^2}{2}\log^2L + t^2c\Bigr) + O(t^3),\ee
implying the anomalous dimension
\be \la{answer1orth}\Delta_q=q(1-q)(t -t^2) +O(t^3),\quad\beta=1.\ee
In view of the long-scale nature of the calculation, one can try to repeat it replacing all momentum sums with integrals. The values of individual computed quantities will change. For instance, one will have $\Pi_2=1$, rather than $\pi^2/12$, which could conceivably destroy the vital cancelation of the $O(k_\m)$ term in (\ref{kinetic term}). However, this does not happen: the prefactor in front of $k_\m$ in the first term of (\ref{term to compute}) also changes so as to preserve the cancelation, and in this way the anomalous dimensions (\ref{answer1unit}) and (\ref{answer1orth}) are found correctly. This is a clear manifestation of the fact that multifractality is insensitive to the discreteness of the model.

\section{Critical ultrametric model on a binary tree}

The critical ultrametric random matrix model (with binary splitting at each level of the hierarchy) was introduced in \cite{FOR09}.
The model itself can not be exactly mapped onto the nonlinear $\sigma$-model. However, by associating with
every site of the underlying lattice $n$ internal degrees of freedom (``orbitals") and coupling them by Gaussian random matrices
in a way respecting the original lattice structure (see e.g. \cite{Zirn} for a more detailed description of the procedure), one can derive an effective nonlinear $\sigma-$model  in the limit $n\to \infty$. This $\sigma$-model is expected to share the critical properties with the original microscopic model. Performing the necessary steps in the ultrametric case, we once again arrive at the non-local $\sigma$-model (\ref{0}) with the kernel $U_{r,r'}$  expressed in terms of the variance matrix of the ultrametric random Hamiltonian $A_{r,r'}=\langle |H_{r,r'}|^2\rangle$ as $U_{r,r'}=(A^{-1})_{r,r'}-\delta_{r,r'}\Bigl(\sum_i A_{i,j}\Bigr)^{-1}$.
Now the system size can only be $L=2^k$, where the integer $k$ is the number of levels in the hierarchical construction. The variance matrix
$A$ is constructed recursively:
\be\fl\quad A_{k=1}=\left( \begin{array}{cc}
\omega & 1\\
1 & \omega
\end{array} \right),\quad\quad  \omega\geq 0, \qquad A_{k}=\left( \begin{array}{cc}
A_{k-1} & 4^{1-k}\mathbf{E}\\
4^{1-k}\mathbf{E} & A_{k-1}
\end{array} \right),\ee
where $\mathbf{E}$ is a square matrix with all entries equal to $1$.

Unlike the power-law case, the value of the coupling constant $t$ in (\ref{0}) is not pre-determined by microscopic parameters, but
can be chosen {\it ad hoc}. Being mainly interested in developing the perturbative treatment of the model, we consider
this coupling to be small. Anticipating the result, it is actually convenient to replace $t$ in (\ref{0}) with the combination
\be\la{ttilda} \w t=\frac{6\log 2}{(\omega+2)^2}\, t,\quad t\ll 1.\ee
In order to diagonalize matrices with ultrametric block structure, it is technically convenient to use the so-called replica Fourier transform (RFT),
as presented in \cite{DG2006}.
The quantities of interest are then written as discrete sums, which can be subsequently expressed in terms of the $q$-digamma function with $q=2$, whose asymptotics are known \cite{AS1972}. For instance, we obtain $\Pi_{r,r}=-(\omega+2-2^{1-k})\sum_{j=1}^k 2^{j-1}\frac{\omega+2 -3\cdot 2^{j-k}}{2-3\cdot 2^j}$, whence $\Pi_{r,r}=\frac{(\omega+2)^2}{6\log 2}\log L +O(1)$. 
Proceeding in this way, we recover in the unitary case the same formulas (\ref{preliminary1unit}) and (\ref{answer1unit}) as in the PLBRM model.
In the orthogonal case, we again need to take care of the contribution from the ``kinetic'' terms (\ref{kinetic term}). 
After a rather tedious calculation we arrive at the final result:
\be\fl I^{(O)}_q\!=\!\frac{(2q\!-\!1)!!}{L^{q-1}}\Bigl(1\!+\! (t\!-\!t^2\frac{13}{6}\log 2)q(q\!-\!1)\log L \!+\!t^2\frac{q^2(q\!-\!1)^2}{2}\log^2L\!+\!t^2c\Bigr)\!+\!O(t^3),\ee
\be\la{answer2orth}\Delta_q=q(1-q)\Bigl(t-t^2\frac{13}{6}\log2\Bigr) +O(t^3),\quad\beta=1.\ee
In contrast with the power-law case, the anomalous dimensions for the ultrametric case could not be obtained by simply replacing sums with integrals.

\section{Non-parabolic corrections to multifractal spectrum in the unitary case}
We have found that in the first two orders the anomalous dimensions $\Delta_q$ are given by a quadratic polynomial in $q$. This means that the multifractal spectrum $f(\alpha)$, obtained from $d_q$ by a Legendre transform \cite{EM08},
is parabolic. The first correction to $\Delta_q$ which induces non-parabolicity of $f(\alpha)$
occurs in the next order $O(t^3)$, where we restrict our calculation to the unitary symmetry class ($\beta=2$) for simplicity.
Continuing the perturbation series, we find in the PLBRM case
\be\fl I^{(U)}_q=C&\frac{q!}{L^{q-1}}\Bigl[1+t\frac{q(q-1)}{2}\Pi_{r,r} +t^2\frac{q^2(q-1)^2}{8}\Pi^2_{r,r} + t^3\frac{q^3(q-1)^3}{48}\Pi^3_{r,r}+\nonumber \\\fl &t^3A(L)\,q^2(q-1)^2+t^3B(L)\,q(q-1) + t^3 c\Bigr] + O(t^4).\ee
In the ultrametric case, the expression is the same up to a change $t\to\w t$. A prefactor $C=1-t^2\frac{q(q-1)(2q-1)}{4}\Pi_2$ has been isolated. One can see that non-parabolic corrections to $\Delta_q$ arise only from the term denoted $A(L)$, which reads $A(L)=-\frac{1}{6}\Pi^3_{r,r}+\frac{1}{4L}\sum_{r_1,r_2}\Pi^3_{r_1,r_2} +\frac{1}{8L}\sum_{r_1,r_2,r_3}U_{r_1,r_2}\Pi^2_{r_3,r_1}\Pi^2_{r_3,r_2}.$
We computed the sums numerically in both models and fitted the result by a polynomial of $\log L$.
The absence of $\log^3L$ and $\log^2L$ terms can be stated with confidence: in our numerical calculations the coefficients in front of these terms were $\lesssim10^{-4}$. These terms are expected to be absent, since their presence would be incompatible with the multifractal scaling. As for the linear term, it was found to be {\it different} in the two models:
\be \fl\la{numerics}A_{\mathrm{plbrm}}(L)= - y_{\mathrm{plbrm}}\log L + c,\quad A_{\mathrm{ultra}}(L)=-\Bigl(\frac{t}{\w t}\Bigr)^3y_{\mathrm{ultra}}\log L + c,\ee
where $\w t/t$ was given in eq. (\ref{ttilda}) and
\be\la{difference} y_{\mathrm{plbrm}}=0.407\pm 0.004,\quad y_{\mathrm{ultra}}=0.486\pm0.008.\ee
The dependence of $A(L)$ on $\omega$ in eq. (\ref{numerics}) in the ultrametric case is natural: it means that $\Delta_q$ depends on $\omega$ only via $t$. We also checked this fact numerically for three values: $\omega=0$, 0.5 and 1. As a result, the non-parabolic part of the anomalous dimension is given, both in PLBRM and ultrametric cases, by
\be \la{non-parabolic} \Delta_q^{\mathrm{non \mhyphen par}}=t^3\,y\, q^2(1-q)^2 +O(t^4),\quad\beta=2,\ee
with the value of $y$ taken from (\ref{difference}). The fact that this value is different for the two models, while in the first two orders their dimensions were identical (eq. (\ref{answer1unit})), demonstrates the announced ``non-universality" of the anomalous exponents beyond the parabolic approximation.

\section*{Conclusion}
The second order corrections to the anomalous dimension for $\beta=1$ (eqq. (\ref{answer2orth}) and (\ref{answer1orth})) appear different in the two models: it is roughly $50\%$ larger in the ultrametric case ($\frac{13}{6}\log2\approx 1.5018$). A somewhat similar situation (identical first order, but different second order corrections) arises in the comparison of the classical Ising model with a power-law interaction and the hierarchical Dyson model\footnote{The authors are grateful to the anonymous referee for bringing this fact to their attention.} \cite{KT1977}. However, since the $q$-dependence of our $\Delta_q$ is parabolic in the first two orders, this difference could be in principle absorbed by a suitable redefinition of the coupling constant in the ultrametric case. To resolve the question of alleged non-equivalence of the two models, and hence of the non-universality of the anomalous dimensions in long-ranged systems, we extended our analysis of the unitary case to the next order of the perturbation series and calculated the first non-parabolic contribution to $\Delta_q$. We observed a difference between the two models (eq. (\ref{difference})), which can no longer be removed by a simple redefinition of couplings.

Instead of a simple scaling law (\ref{formula number 1}), $I_q$ can in principle behave as a sum of several different powers of $L$, the highest of which dominates in the limit $L\to \infty$ and gives the multifractal spectrum. Nevertheless, up to the second order in the coupling constant our results are consistent with the single-power scaling (\ref{formula number 1}). This is true for any non-local $\sigma$-model and not only for the two examples considered here.

IR and AO acknowledge support from the Engineering and Physical Sciences Research Council [grant number EP/G055769/1].

\end{document}